\documentstyle[aps,pra]{revtex}
\begin{document}
\draft

\title{
Nonlocally looking equations can make nonlinear quantum dynamics
local
}
\author{Marek Czachor}
\address{
Katedra Fizyki Teoretycznej i Metod Matematycznych\\
 Politechnika Gda\'{n}ska,
ul. Narutowicza 11/12, 80-952 Gda\'{n}sk, Poland\\
and \\
Arnold Sommerfeld Institute for Mathematical Physics\\
Technical University of Clausthal, D-38678 Clausthal-Zellerfeld,
Germany
}
\maketitle
\begin{abstract}
A general method for extending a non-dissipative 
nonlinear Schr\"odinger and
Liouville-von Neumann 1-particle dynamics to an arbitrary number
of particles is described. It is shown at a general level that the
dynamics so obtained is completely separable, which is 
the strongest condition
one can impose on dynamics of composite systems. It requires
that for all initial states (entangled or not) 
a subsystem not only cannot be influenced by any
action undertaken by an observer in a separated system (strong
separability), but additionally that the self-consistency
condition ${\rm Tr\,}_2\circ \phi^t_{1+2}=\phi^t_{1}\circ {\rm Tr\,}_2$
is fulfilled. 
It is shown that a correct extension
to $N$ particles involves integro-differential equations which,
in spite of their nonlocal appearance, make the theory fully local.
As a consequence a much larger class of
nonlinearities satisfying the complete separability condition 
is allowed than has been assumed so far. In particular all 
nonlinearities of the form $F(|\psi(x)|)$ are acceptable. 
This shows that the locality condition does not single out
logarithmic or 1-homeogeneous nonlinearities.
\end{abstract}

\pacs{PACS numbers: 03.65.Bz, 03.75.Fi, 05.30.-d, 11.10.Lm}
\vskip1pc

\section{Introduction}

In spite of the linearity of the Schr\"odinger and Liouville-von
Neumann equations, nonlinearly evolving states are 
encountered in quantum mechanics quite often. Typically this is
a result of approximations used in a description of collective
phenomena (cf. nonlinear interferences in a Bose-Einstein
condensate \cite{BEC1})
but no proof has been given so far that it is not the
{\it linearity\/} that is a result of some approximation. 
This obvious observation motivated many authors to either look
for nonlinear extensions of quantum mechanics, or to try to find
an argument against a nonlinear evolution at a fundamental
level. 

A privileged role in both cases was played by a separability
condition. Apparently the first use of the condition can be
found in \cite{BBM}. The authors required that a nonlinearity
must allow two separated, noninteracting and uncorrelated
subsystems to evolve independently of each other, and this (plus
some additional assumptions) led them to the nonlinear term $\ln
\big(|\psi(x)|\big) \psi(x)$. Similar argument (with
different additional assumptions) led Haag and Bannier \cite{HB}
and Weinberg \cite{W} to the 1-homogeneity condition (i.e. $\psi(x)$ 
and $\lambda\psi(x)$ should satisfy the same equation). An
important element of these works was the assumption that the
systems are uncorrelated, that is their wave function is a
product one. Equations that have such a locality property may be
called {\it weakly separable\/}~\cite{weak-strong}. 
It was
shown later by Gisin and others \cite{W,G1,G2,MCfpl} 
that a weakly separable equation
may still violate causality if an initial state of the composite
system is entangled. A slightly
different in spirit analysis of separability conditions was
given in \cite{GS} where it was argued 
that an extension from  1 to $N$ particles may
lead to new effects that become visible for $N>N_0$, where $N_0$
is a parameter characterizing a given hierarchy of theories.

It was originally conjectured that these properties 
refute any deterministic (i.e.
non-stochastic) nonlinear generalization of quantum mechanics.
That this is not the case was shown for Weinberg-type quantum
mechanics by Polchinski \cite{P} (for pure states) and Jordan
\cite{J} (for density matrices). These results were subsequently
generalized to a more general class of theories (Lie-Nambu
dynamics) by myself in \cite{MCpla}. Therefore, contrary to a
rather common belief, there exists a
class of nonlinear generalizations of quantum mechanics that
does not lead to the locality problems for general (pure-entangled
and general-mixed) states. Such theories can be called {\it
strongly separable\/}. The strongly separable theories
considered so far involve equations which are Hamiltonian, which
means
there exists a Hamiltonian {\it function\/} that generates the
dynamics. Actually, the solution of the problem proposed by
Polchinski in \cite{P} was based on an appropriate choice of
this function. In addition, the Polchinski approach was
formulated within a finite-dimensional framework. 
Still, there exists an interesting class of {\it non-Hamiltonian\/} 
and infinite-dimensional nonlinear equations. For example, it is
known that all Doebner-Goldin equations that are Hamiltonian are
linearizable \cite{linearizable} and therefore their strong separability
may not be very interesting. Quite recently the problem of
strong separability of non-Hamiltonian Doebner-Goldin equations
was addressed in \cite{LN}.
Using the typical extension of the dynamics from 1 to $N$
particles the authors showed that at $t=0$ the time derivatives
(up to the 3rd order) 
of a 1-particle probability density in position space do not depend on
the potential applied to the other particle if the equation is
Galilean-covariant. This {\it suggests\/} that  
Galilean covariance may lead to strong locality in
nonrelativistic domain. The
result agrees with computer algebraic tests undertaken by Werner
\cite{Werner} who chose oscillator-type potentials and wave
functions satisfying certain Gaussian Ansatz. The approach
chosen in those works is essentially a third order perturbation
theory applied to diagonal elements of reduced density
matrices. 
A technical detail that did not allow these authors to find a
general non-perturbative solution was that they investigated a dynamics of
diagonal elements of a
non-pure reduced density matrix (typical of entangled states) 
but the formalism they used was devised
for state vectors and, hence, they had no control over the
behavior of non-pure states and off-diagonal elements
of density matrices~\cite{dm}.  

In this paper I will give a general and model independent 
solution of the problem. I
will generalize to non-Hamiltonian systems the technique that
proved useful in the context of density matrix equations,
namely the triple-bracket formalism \cite{MCpla,diss}. I will
then show how to 
extend a 1-particle dynamics to $N$-particle systems in a strongly
separable way for a large class of both Hamiltonian and
non-Hamiltonian Schr\"odinger  non-dissipative equations. 
As opposed to basically all previous papers dealing with pure
state nonlinear dynamics (the exception is \cite{P}, but it
deals with finite dimensional systems) I will not {\it guess\/}
the ``obvious" form of the extension but will {\it derive\/} it.
This somewhat 
long path beginning with pure states, then extension via
mixtures, and again back to pure states, will be rewarded because the
$N$-particle form we will get will not be the one one might expect. 
Its paricularly interesting feature is the fact that the
equations are {\it integro-differential\/}. This feature is
implicitly present also in the Polchinski-Jordan formalism but
is hidden behind the finite dimensional convention where the
integrals do not explicitly show up. 
The approach is applicable to {\it all\/} non-dissipative
Doebner-Goldin equations \cite{DGpra,DG}, 
the equations discussed in \cite{BBM,HB,W,Twarock},
as well as a large class of nonlinear
Schr\"odinger equations that were not discussed so far.

I will prove simultanously a much stronger result.
By the very construction the formalism is applicable to those
Liouville-von Neumann nonlinear equations that reduce to the
corresponding nonlinear Schr\"odinger dynamics on pure states. All these
equations will be shown to be not only {\it strongly separable\/} but 
also {\it completely separable\/}. By the latter I mean a
strongly separable dynamics which additionally satisfies the
self-consistency condition 
\begin{eqnarray}
{\rm Tr\,}_2\circ \phi^t_{1+2}=\phi^t_{1}\circ {\rm Tr\,}_2,\label{self}
\end{eqnarray}
where $\phi^t_{1+2}$ and $\phi^t_{1}$ denote the dynamics of the
composite system and the subsystem, respectively, and the
partial trace ${\rm Tr\,}_2$ is
a map that reduces the dynamics from the large system to the
subsystem. Condition (\ref{self}), which is independent of strong
separability, was recently pointed out as an
important ingredient of nonlinear dynamics that may be regarded as
completely positive \cite{MCprl}. We will also see that not only
the probability densities in position space but also the
off-diagonal elements of reduced density matrices are
independedent of details of interaction in the remote systems. 

\section{Example: Haag-Bannier equation and its
almost-Lie-Poisson form}

The general convention I will use was elaborated in detail in
\cite{MCpla,MCMK} but to make this work self-contained let us
first explain the general scheme on a concrete example, the
Haag-Bannier equation \cite{HB}. This equation is of the
Doebner-Goldin type, that is contains a 1-homogeneous nonlinear
term with derivatives, but is simpler:
\begin{eqnarray}
i\hbar\partial_t\psi(a) &=&
\Bigl(
-\frac{\hbar^2}{2m}\Delta_a + V(a)
\Bigr)\psi(a)
+
\vec A(a)
\frac{\bar \psi(a)\vec \nabla_a \psi(a) -\psi(a)\vec \nabla_a
\bar \psi(a)}{2i|\psi(a)|^2}
\psi(a).
\end{eqnarray}
Its Liouville-von Neumann counterpart is 
\begin{eqnarray}
i\hbar\partial_t \rho(a,a')&=&
\Bigl(
-\frac{\hbar^2}{2m}\Delta_{a} + V(a)
\Bigr)\rho(a,a')
+
\vec A(a)
\frac{\int dy\,\delta(a-y)\vec \nabla_{a}[ \rho(a,y)
- \rho(y,a)]}
{2i\rho(a,a)} 
\rho(a,a')\nonumber\\
&\phantom{=}&-
\Bigl(
-\frac{\hbar^2}{2m}\Delta_{a'} + V({a'})
\Bigr)\rho(a,a')
-
\vec A(a')
\frac{\int dy\,\delta(a'-y)\vec \nabla_{a'}[ \rho(a',y)
- \rho(y,a')]}
{2i\rho(a',a')} 
\rho(a,a'),\label{LvN}
\end{eqnarray}
where $\rho(a,a')=\overline{\rho(a',a)}$, and $f(a)=\rho(a,a)\geq
0$ is
$d^3a$-integrable~\cite{what} together with all its natural
powers $f(a)^n$. For
$\rho(a,a')=\psi(a)\overline{\psi(a')}$ (\ref{LvN}) reduces to
the Schr\"odinger-Haag-Bannier dynamics. 
Using the composite index convention described in
\cite{MCpla} we can write the equation in a form which is
compact and simplifies general calculations.
Denote $\rho_a=\rho(a,a')$ and 
\begin{eqnarray}
H^a(\rho)&=&
H(a',a)=
K(a',a) + V(a)\delta(a-a')
+
\vec A(a)
\frac{\int dy\,\delta(a-y)\vec \nabla_{a}[ \rho(a,y)
- \rho(y,a)]}
{2i\rho(a,a)} \delta(a-a').
\end{eqnarray}
The kinetic kernel satisfies 
\begin{eqnarray}
\int dy\,K(a,y)\rho(y,a')=-\frac{\hbar^2}{2m}\Delta_{a}\rho(a,a'),
\end{eqnarray}
and $K(a,b)=\overline{K(b,a)}$.
All the integrals are in $\bbox R^3$ i.e. $da=d^3a$, etc., and
the ``summmation convention" is applied at the composite index
level (two repeated indices are inegrated). 
There is no conflict of notation here because the
composite indices are always in their upper or lower positions
whereas the 3-dimensional coordinates $a$ are arguments of
functions or distributions. (Notice that the composite indices
correspond always to pairs of primed and unprimed 3-dimensional
coordinates.)  
The indices are raised by the metric $g^{ab}$ 
working as follows
\begin{eqnarray}
g^{ab}\rho_b
&=&
\int db db'\delta(a-b')\delta(b-a')\rho(b,b')=\rho(a',a)=\rho^a.
\nonumber
\end{eqnarray}
So if $\rho_a=\rho(a,a')$ then $\rho^a=\rho(a',a)$. 
To lower an index one uses obvious inverse formulas \cite{MCpla}. 

The Liouville-von Neumann equation can be written in a
triple-bracket-type form \cite{MCpla,Mor,MCijtp}
\begin{eqnarray}
i\hbar\partial_t\rho_f &=&
\int db  db' dc  dc'
\Big(
\underbrace{
\delta(f-b')\delta(b-c')\delta(c-f')
-
\delta(f-c')\delta(b-f')\delta(c-b')
}_{\Omega_{fbc}}\Big)
\underbrace{H(b',b)}_{H^b} \underbrace{\rho(c',c)}_{\rho^c}\\
&=&
\Omega_{fbc}H^b\rho^c=
\Omega_{abc}
\frac{\delta \rho_f}{\delta \rho_a}H^b\rho^c.\label{triple}
\end{eqnarray}
Let us note that if 
\begin{eqnarray}
H^b=\frac{\delta H}{\delta \rho_b}\label{H}
\end{eqnarray}
then (\ref{triple}) describes a Lie-Poisson dynamics of a density
matrix \cite{Bona}. If no Hamiltonian function $H$ satisfying
(\ref{H}) exists, the dynamics will be called 
almost-Lie-Poisson. The generic case discussed in this Letter is
almost-Lie-Poisson. The Weinberg-type and mean-field dynamics
discussed in \cite{J,Bona} are Lie-Poisson. 
The discussion presented here applies to a general
almost-Lie-Poisson dynamics where the structure constants have
the following general form \cite{MCpla}
\begin{eqnarray}
\Omega{_{abc}}&=&
I_{\alpha\beta'}
I_{\beta\gamma'}
I_{\gamma\alpha'}
-
I_{\alpha\gamma'}
I_{\beta\alpha'}
I_{\gamma\beta'}\label{O_}\\
\Omega{^{abc}}&=&
-\omega^{\alpha\beta'}
\omega^{\beta\gamma'}
\omega^{\gamma\alpha'}
+
\omega^{\alpha\gamma'}
\omega^{\beta\alpha'}
\omega^{\gamma\beta'}\label{O^}
\end{eqnarray}
where $\omega^{\alpha\alpha'}=\omega^a$ and $I_{\alpha\alpha'}=I_a$ are,
respectively, the symplectic form and the Poisson tensor
corresponding to the pure state equation. Here
$\omega^a=\delta(a-a')$, $I_a=\delta(a-a')$ but (\ref{O_}) and 
(\ref{O^}) are valid also for other Hilbert spaces and equations
(cf. \cite{MCpla,MCMK}).

\section{2-particle extension}

Consider now a 2-particle system described by the density matrix
which, depending on whether the state is pure or general, is in
either of the two forms
\begin{eqnarray}
\rho_a=\rho_{a_1a_2}&=&\rho(a_1,a_2,a_1',a_2')
\label{12a}\\
&{\stackrel{\rm pure}{=}}&
\Psi(a_1,a_2)\overline{\Psi(a_1',a_2')}.
\label{12b}
\end{eqnarray}
The reduced density matrices are 
\begin{eqnarray}
\rho^{I}_{a_1}&=&\int dy\,\rho(a_1,y,a_1',y)=
\omega^{a_2}\rho_{a_1a_2}
\\
\rho^{II}_{a_2}&=&\int dy\,\rho(y,a_2,y,a_2')=
\omega^{a_1}\rho_{a_1a_2}.
\end{eqnarray}
The 2-particle Liouville-von Neumann-Haag-Bannier equation is 
given by (\ref{triple}) but with all indices doubled like in
(\ref{12a}), (\ref{12b})
with the 2-particle structure constants given explicitly by
\begin{eqnarray}
\Omega_{abc}^{(2)}
&=&
\delta(a_1-b_1')\delta(b_1-c_1')\delta(c_1-a_1')\nonumber\\
&\phantom{=}&\times
\delta(a_2-b_2')\delta(b_2-c_2')\delta(c_2-a_2')\nonumber\\
&-&
\delta(a_1-c_1')\delta(b_1-a_1')\delta(c_1-b_1')\nonumber\\
&\phantom{=}&\times
\delta(a_2-c_2')\delta(b_2-a_2')\delta(c_2-b_2').
\end{eqnarray}
The crucial element of the whole construction is the
nonlinear extension of the 1-particle Hamiltonian operator
kernel. We {\it define\/} it as
\begin{eqnarray}
H^b(\rho)
&=&
H_{I}^{d_1}(\rho^{I})\frac{\delta\rho^{I}_{d_1}}
{\delta\rho_{b}}
+
H_{II}^{d_2}(\rho^{II})\frac{\delta\rho^{II}_{d_2}}
{\delta\rho_{b}}\label{H1+H2}\\
&=&
H_{I}^{b_1}(\rho^{I})\omega^{b_2}
+
\omega^{b_1}
H_{II}^{b_2}(\rho^{II}),
\end{eqnarray}
with $b=b_1b_2$. 
(\ref{H1+H2}) is a nonlinear functional generalization of the
well known linear recipe $H_{1+2}=H_1\otimes \bbox 1 + 
\bbox 1\otimes H_2$ and reduces to the formulas arising from the
triple-bracket formalism if Hamiltonian functions exist.
The whole 2-particle equation can be written as
follows 
\begin{eqnarray}
i\hbar\partial_t\rho_f
&=&
\Omega_{abc}^{(2)}
\frac{\delta \rho_f}{\delta \rho_a}
\Bigg(
H_{I}^{d_1}(\rho^{I})\frac{\delta\rho^{I}_{d_1}}
{\delta\rho_{b}}
+
H_{II}^{d_2}(\rho^{II})\frac{\delta\rho^{II}_{d_2}}
{\delta\rho_{b}}\Bigg)\rho^c\label{eq}
\end{eqnarray}
where $\rho_f=\rho_{f_1f_2}$ is the 2-particle density matrix. 

Now comes the important general result. Let us perform a partial
trace i.e. contract both sides of (\ref{eq}) with
$\omega^{f_2}$. Since $\rho^{I}_{f_1}=\omega^{f_2}\rho_{f_1f_2}=
I_{f_2}\rho{_{f_1}}^{f_2}$
and $\omega^{f_2}=\delta(f_2-f_2')$ is $t$- and
$\rho$-independent  
\begin{eqnarray}
&{}&i\hbar\partial_t\rho^{I}_{f_1}\nonumber\\
&{}&\phantom{=} =
\Omega_{abc}^{(2)}
\frac{\delta \rho^{I}_{f_1}}{\delta \rho_a}
\Bigg(
H_{I}^{d_1}(\rho^{I})\frac{\delta\rho^{I}_{d_1}}
{\delta\rho_{b}}
+
H_{II}^{d_2}(\rho^{II})\frac{\delta\rho^{II}_{d_2}}
{\delta\rho_{b}}\Bigg)\rho^c\label{1}\\
&{}&\phantom{=} =
\Omega_{abc}^{(2)}
\frac{\delta \rho^{I}_{f_1}}{\delta \rho_a}
H_{I}^{d_1}(\rho^{I})\frac{\delta\rho^{I}_{d_1}}
{\delta\rho_{b}}\rho^c\label{2}\\
&{}&\phantom{=} =
\Omega_{f_1d_1c_1}^{(1)}
H_{I}^{d_1}(\rho^{I})
\rho^{I\,c_1}.\label{4}
\end{eqnarray}
where the transition from (\ref{1}) to (\ref{2}) is a
consequence of 
\begin{eqnarray}
\Omega{^{(N)}}{_{abc}}
\frac{\delta \rho{^I}_{d}}{\delta \rho_{a}}
\frac{\delta \rho{^{II}}_{e}}{\delta \rho_{b}}=0\label{lem1}
\end{eqnarray}
holding for all $N$-particle structure constants and reduced
density matrices of non-overlapping subsystems (for the proof of
(\ref{lem1}) see
Lemma~1 in \cite{MCpla}). 

Let me summarize what has happened until now: Using a correct
extension of a 1-particle nonlinear Hamiltonian operator and the
general property of triple brackets we have reduced a 2-particle
equation for a 2-particle density matrix to a 1-particle
equation which involves {\it only\/} the quantities which are
intrinsic to this subsystem. All elements depending on the other
subsystem have simply vanished. We have obtained this by
performing only one operation --- the partial trace over the
``external" subsystem. 

Therefore the reduced density matrix in $I$ does not depend on
details of interaction in the separated system $II$ and the
dynamics is strongly separable. Actually,
we have simultaneously obtained more. Indeed, we do not assume
that we take the partial trace at ``$t=0$". Therefore we can
trace out the external system at any time and the reduced
dynamics is indistinguishable from a dynamics defined entirely in terms
of the subsystem and starting from the initial condition 
$\rho_1(0)=\rho^I(0)={\rm Tr\,}_2\rho(0)$ where $\rho(0)$ is the
initial condition for the large system. It proves that the
dynamics so constructed is {\it completely separable\/}. 

The result we have obtained is completely general and works for
all Schr\"odinger equations whose nonlinear Hamiltonian operator
kernels allow to write them in terms of 1-particle density matrices. 
Putting it differently, the construction works correctly 
if a
given Schr\"odinger equation allows for an extension to an
almost-Lie-Poisson Liouville-von Neumann equation. 
The example of the Haag-Bannier equation served only as a means of
focusing our attention and making the discussion less abstract. 

So how does the 2-particle equation look explicitly? Beginning
again with the general formula we can obtain immediately its
model independent form just by performing abstract
operations on the composite indices. We get 
\begin{eqnarray}
i\hbar\partial_t\rho_{a}
&=&
\Omega_{abc}^{(2)}
\Bigg(
H_{I}^{b_1}(\rho^{I})\omega^{b_2}
+
\omega^{b_1}H_{II}^{b_2}(\rho^{II})\Bigg)\rho^c\nonumber\\
&=&
\Omega_{a_1b_1c_1}^{(1)}
H_{I}^{b_1}(\rho^{I})\rho^{c_1}{_{a_2}}
+
\Omega_{a_2b_2c_2}^{(1)}
H_{II}^{b_2}(\rho^{II})\rho{_{a_1}}^{c_2}.\nonumber
\end{eqnarray}
Returning for the sake of completeness to the Haag-Bannier case
we can write it as 
\begin{eqnarray}
{}&{}&i\hbar\partial_t \rho(a_1,a_2,a_1',a_2')\nonumber\\
&{}&=
\Bigg[
\Bigl(
-\frac{\hbar^2}{2m}\Big(\Delta_{a_1} 
+ \Delta_{a_2} -\Delta_{a_1'}- \Delta_{a_2'}\Big)
+ V_1(a_1) +V_2(a_2) - V_1(a_1') - V_2(a_2')
\nonumber\\
&{}&
+\vec A_1(a_1)
\frac{\int dy_1\,\delta(a_1-y_1)\vec \nabla_{a_1}
[\int dz\, \rho(a_1,z,y_1,z)
- \int dz\,\rho(y_1,z,a_1,z)]}
{2i\int dz\,\rho(a_1,z,a_1,z)} 
\nonumber\\
&{}&
+
\vec A_2(a_2)
\frac{\int dy_2\,\delta(a_2-y_2)\vec \nabla_{a_2}
[\int dz\, \rho(z,a_2,z,y_2)
- \int dz\,\rho(z,y_2,z,a_2)]}
{2i\int dz\,\rho(z,a_2,z,a_2)} 
\nonumber\\
&{}&-
\vec A_1(a_1')
\frac{\int dy_1\,\delta(a_1'-y_1)\vec \nabla_{a_1'}
[\int dz\, \rho(a_1',z,y_1,z)
- \int dz\,\rho(y_1,z,a_1',z)]}
{2i\int dz\,\rho(a_1',z,a_1',z)} 
\nonumber\\
&{}&-
\vec A_2(a_2')
\frac{\int dy_2\,\delta(a_2'-y_2)\vec \nabla_{a_2'}
[\int dz\, \rho(z,a_2',z,y_2)
- \int dz\,\rho(z,y_2,z,a_2')]}
{2i\int dz\,\rho(z,a_2',z,a_2')} 
\Bigg]
\rho(a_1,a_2,a_1',a_2').
\end{eqnarray}
Its {\it pure state\/} 2-particle counterpart is 
\begin{eqnarray}
i\hbar\partial_t \Psi(a_1,a_2)
&=&
\Bigg[
-\frac{\hbar^2}{2m}\big(\Delta_{a_1}+\Delta_{a_2}\big) 
+ V_1(a_1) + V_2(a_2)\nonumber\\
&\phantom{=}&
+
\vec A_1(a_1)
\frac{\int dz\big[\overline{\Psi(a_1,z)} \vec \nabla_{a_1}\Psi(a_1,z)
- \Psi(a_1,z)\vec \nabla_{a_1}\overline{\Psi(a_1,z)}\big]}
{2i\int dz|\Psi(a_1,z)|^2} 
\nonumber\\
&\phantom{=}&+
\vec A_2(a_2)
\frac{
\int dz\, \big[\overline{\Psi(z,a_2)}\nabla_{a_2}\Psi(z,a_2)
- \Psi(z,a_2)\nabla_{a_2}\overline{\Psi(z,a_2)}\big]}
{2i\int dz |\Psi(z,a_2)|^2} 
\Bigg]
\Psi(a_1,a_2).
\end{eqnarray}
This equation has several interesting features. The Hamiltonian
operator  obviously
reduces to the sum of ordinary 1-particle terms, involving no
integrals, if $\Psi$ is a 
product state. What makes it unusual is the presence of
the integrals. Typically it is said that such equations
should not be taken into account because they are {\it
nonlocal\/}. They indeed appear nonlocal but a closer look shows
that it is in fact just the opposite: The currents and probability
densities are the local 1-particle ones. Therefore it is the lack of
{\it appropriate\/} inegrals that makes typical 2-particle
equations nonlocal. 
The equation considered in \cite{LN} involved no such integrals
and this led to difficulties.
An extension of the above results from 2 to $N$ particles is
immediate so explicit formulas will not be given. 

\section{Further examples}

Let me now list some of nonlinear Hamiltonian operators 
that have been considered in
literature and which admit the completely separable extension to
an arbitrary number of particles in arbitrary entangled and mixed states.
The general rule is that an extension to $N$ particles will be
given in one of these forms but instead of $\rho(x,x')$ a reduced
1-particle density matrix should be placed. If the $N$-particle state is
pure then the reduced density matrix is a functional of the pure
state which involves $N-1$ integrations. This matrix is
subsequently put into a suitable place in the Schr\"odinger
equation. Similarly it can be put into a Liouville-von Neumann
equation if the state is more general.

\medskip
\noindent
a) {\it ``Nonlinear Schr\"odinger"\/}
\begin{eqnarray}
|\psi(x)|^2\to \rho(x,x)\nonumber
\end{eqnarray}
b) {\it Bia{\l}ynicki-Birula--Mycielski\/}
\begin{eqnarray}
\ln\big(|\psi(x)|^2\big)\to \ln\rho(x,x)\nonumber
\end{eqnarray}
Obviously in the same way one can treat any equation with
nonlinearities given by some function $F(|\psi(x)|)$\cite{problem}.

\noindent 
c) {\it Doebner--Goldin\/}
\begin{eqnarray}
R_1:&{}&
\frac{1}{2i}
\frac{
\bar\psi(x)\Delta_x \psi(x)
-
\psi(x)\Delta_x \bar\psi(x)}
{|\psi(x)|^2}\nonumber\\
&{}&\to
\frac{1}{2i}
\frac{\int dz\delta(x-z)
\Delta_x \big[\rho(x,z)
-\rho(z,x)\big]}
{\rho(x,x)},\nonumber\\
R_2:&{}&
\frac{\Delta_x|\psi(x)|^2}{|\psi(x)|^2}
\to
\frac{\Delta_x\rho(x,x)}{\rho(x,x)}\nonumber\\
R_3:&{}&
\frac{1}{(2i)^2}
\frac{
\bigl[
\bar \psi(x)\vec \nabla_x \psi(x)
-
\psi(x)\vec \nabla_x \bar \psi(x) \bigr]^2}
{|\psi(x)|^4}\nonumber\\
&{}&
\to
\frac{1}{(2i)^2}
\frac{\big(
\int dz\delta(x-z)
\vec \nabla_x \big[\rho(x,z)
-\rho(z,x)\big]\big)^2}
{\rho(x,x)^2}\nonumber\\
R_4:&{}&
\frac{1}{2i}
\frac{
\bigl[
\bar \psi(x)\vec \nabla_x \psi(x)
-
\psi(x)\vec \nabla_x\bar \psi(x)
\bigr]\cdot\vec \nabla_x |\psi(x)|^2}
{|\psi(x)|^4}\nonumber\\
&{}&\to
\frac{1}{2i}
\frac{\int dz\delta(x-z)
\vec \nabla_x \big[\rho(x,z)
-\rho(z,x)\big]\cdot \vec \nabla_x\rho(x,x)}
{\rho(x,x)^2},\nonumber\\
R_5:&{}&
\frac{
\bigl[
\vec \nabla_x |\psi(x)|^2\bigr]^2}
{|\psi(x)|^4}
\to
\frac{
\bigl[
\vec \nabla_x \rho(x,x)\bigr]^2}
{\rho(x,x)^2}\nonumber
\end{eqnarray}
d) {\it Twarock\/} on $S^1$\cite{Twarock}
\begin{eqnarray}
{}&{}&
\frac{\psi(x)'' \overline{\psi(x)'}
-
\overline{\psi(x)''}\psi(x)'}
{\psi(x) \overline{\psi(x)'}
-
\overline{\psi(x)} \psi(x)'}\nonumber\\
&{}&\to
\frac{
[\int dy\delta(x-y)\partial_x^2\rho(x,y)]
[\int dz\delta(x-z)\partial_x^2\rho(z,x)] - c.c.}
{\rho(x,x)\int dy\delta(x-y)\partial_x\rho(x,y) - c.c.}\nonumber
\end{eqnarray}
e) $(n,n)$-{\it homogeneous nonlinearities\/}. Denote by
$D$ a differential operator involving
arbitrary mixed partial derivatives up to order $k$. 
Consider a  real function
$F(\psi)=F\big(D\psi(x)\big)$, 
$(n,n)$-homogeneous i.e. satisfying $F(\lambda\psi)=\lambda^n
\bar \lambda^n F(\psi)$. We first write 
\begin{eqnarray}
F\big(D\psi(x)\big)=
\frac{F\big(\overline{\psi(x)}D\psi(x)\big)}
{|\psi(x)|^{2n}}\nonumber
\end{eqnarray}
and then apply the tricks used for the Haag--Bannier, 
Doebner--Goldin and Twarock
terms. Obviously any reasonable function of such
$(n,n)$-homogeneous expressions with different $n$'s is
acceptable as well.

\section{Extension in stages}

The formalism proposed above allows to extend a 1-particle
dynamics to $N$ particles. We will now show that it allows for a
more general kind of extension: From 1 {\it system\/} to $N$ systems.
The procedure is self-consistent in the sense that one can first
produce several composite systems from the 1-particle ones, and then
combine them into a single overall composite system.
Alternatively, one can produce the final system without the
intermediate stages, directly by
extension from single particles. 

Consider the $N$-particle extension 
\begin{eqnarray}
H_{A}^b(\rho)
&=&
H_{A}^{b_1\dots b_N}(\rho)\nonumber\\
&=&
H_{(1)}^{d_1}(\rho^{(1)})\frac{\delta\rho^{(1)}_{d_1}}
{\delta\rho_{b_1\dots b_N}}
+
H_{(2)}^{d_2}(\rho^{(2)})\frac{\delta\rho^{(2)}_{d_2}}
{\delta\rho_{b_1\dots b_N}}
+\dots
+
H_{(N)}^{d_N}(\rho^{(N)})\frac{\delta\rho^{(N)}_{d_N}}
{\delta\rho_{b_1\dots b_N}}\\
&=&
H_{(1)}^{b_1}(\rho^{(1)})
\omega^{b_2}\dots \omega^{b_N}
+
\omega^{b_1}H_{(2)}^{b_2}(\rho^{(2)})
\omega^{b_3}\dots \omega^{b_N}
+\dots
+
\omega^{b_1}\dots \omega^{b_{N-1}}
H_{(N)}^{b_N}(\rho^{(N)}).
\end{eqnarray}                                                      
The Hamiltonian operator (kernel) $H_{A}^b(\rho)$ describes a
composite system, labelled $A$, which consists of $N$ particles
that do not interact with one another. Consider now another
system, labelled $B$, consisting of $M$ particles. Its
Hamiltonian operator is 
\begin{eqnarray}
H_{B}^b(\rho)
&=&
H_{B}^{b_{N+1}\dots b_{N+M}}(\rho)\nonumber\\
&=&
H_{(N+1)}^{d_{N+1}}(\rho^{(N+1)})\frac{\delta\rho^{(N+1)}_{d_{N+1}}}
{\delta\rho_{b_{N+1}\dots b_{N+M}}}
+
H_{(N+2)}^{d_{N+2}}(\rho^{(N+2)})\frac{\delta\rho^{(N+2)}_{d_{N+2}}}
{\delta\rho_{b_{N+1}\dots b_{N+M}}}
+\dots
+
H_{(N+M)}^{d_{N+M}}(\rho^{(N+M)})\frac{\delta\rho^{(N+M)}_{d_{N+M}}}
{\delta\rho_{b_{N+1}\dots b_{N+M}}}\\
&=&
H_{(N+1)}^{b_{N+1}}(\rho^{(N+1)})
\omega^{b_{N+2}}\dots \omega^{b_{N+M}}
+
\omega^{b_{N+1}}
H_{(N+2)}^{b_{N+2}}(\rho^{(N+2)})
\omega^{b_{N+3}}\dots \omega^{b_{N+M}}
\nonumber\\
&\phantom =&
+\dots
+
\omega^{b_{N+1}}\dots \omega^{b_{N+M-1}}
H_{(N+M)}^{b_{N+M}}(\rho^{(N+M)})
\end{eqnarray}
The $(N+M)$-particle extension can be obtained in two stages
\begin{eqnarray}
H_{A+B}^b(\rho)
&=&
H_{A+B}^{b_{1}\dots b_{N+M}}(\rho)\nonumber\\
&=&
H_{A}^{d_{1}\dots d_{N}}
(\rho^A)
\frac{\delta\rho^{A}_{d_{1}\dots d_{N}}}
{\delta\rho_{b_{1}\dots b_{N+M}}}
+
H_{B}^{d_{N+1}\dots d_{N+M}}(\rho^{B})
\frac{\delta\rho^{B}_{d_{N+1}\dots d_{N+M}}}
{\delta\rho_{b_{1}\dots b_{N+M}}}\nonumber\\
&=&
H_{A}^{b_{1}\dots b_{N}}
(\rho^A)
\omega^{b_{N+1}}\dots \omega^{b_{N+M}}
+
\omega^{b_{1}}\dots \omega^{b_{N}}
H_{B}^{b_{N+1}\dots b_{N+M}}(\rho^{B})\nonumber\\
&=&
\Big(
H_{(1)}^{b_1}(\rho^{(1)})
\omega^{b_2}\dots \omega^{b_N}
+\dots
+
\omega^{b_1}\dots \omega^{b_{N-1}}
H_{(N)}^{b_N}(\rho^{(N)})
\Big)
\omega^{b_{N+1}}\dots \omega^{b_{N+M}}\nonumber\\
&\phantom =&
+
\omega^{b_{1}}\dots \omega^{b_{N}}
\Big(
H_{(N+1)}^{b_{N+1}}(\rho^{(N+1)})
\omega^{b_{N+2}}\dots \omega^{b_{N+M}}
+\dots
+
\omega^{b_{N+1}}\dots \omega^{b_{N+M-1}}
H_{(N+M)}^{b_{N+M}}(\rho^{(N+M)})
\Big)
\end{eqnarray}
which could be derived also directly from the (N+M)-particle
extension of the 1-particle dynamics.

\section{Is nonlinear quantum mechanics nonlocal?}

N.~Gisin in his nowadays classic paper \cite{G1} argued
that {\it any\/} nonlinear and non-stochastic dynamics
necessarily leads to faster-than-light communication via
EPR-type correlations. 
The argument is based (implicitly) on the additional assumption
that a reduced density matrix (of, say, Alice) evolves by means
of an independent 
dynamics of its pure-state components. 
(Let us note that this was essentially the main question discussed by
Haag and Bannier \cite{HB} in the context of the nonlinear, convex scheme
proposed by Mielnik \cite{Mielnik}.)
This viewpoint is
suggested by the fact that what one starts with is a nonlinear
(Schr\"odinger-type) dynamics of pure states. But in the EPR
case the pure state is a 2-particle one and one has to start
with a 2-particle nonlinear Schr\"odinger equation. Assume this
equation is given (e.g. our Haag-Bannier equation) and its
2-particle pure-state solution $|\psi\rangle$ has been found. To
have the Gisin effect one has to find at least one observable of
the form 
\begin{eqnarray}
\langle\hat A\rangle=
\langle\psi|\hat A\otimes 1|\psi\rangle={\rm
Tr}_A\rho_A\hat A\label{alice}
\end{eqnarray}
describing the
Alice subsystem, which depends on a parameter controlled by Bob.
Here 
$\rho_A={\rm Tr}_B|\psi\rangle\langle\psi|$ is the reduced
density matrix of the Alice subsystem. 
Assume that the only actions Bob can undertake are reducible to the
modifications of the parameters of the Hamiltonian corresponding
to ``his" particle. (In the case contemplated by Gisin Bob would
rotate his Stern-Gerlach device; this is equivalent to modifying
the magnetic term in the corresponding interaction Hamiltonian.)
(\ref{alice}) shows that Bob can influence $\langle\hat
A\rangle$ if and only if the reduced density matrix $\rho_A$
depends on the parameter he controls. This is true for the
Weinberg theory \cite{MCfpl} but will {\it never\/} happen for
{\it any\/} equation we have discussed. It follows that the
Gisin effect applies to a limited class of theories. 

The additional assumption that leads to the
Gisin phenomenon is the following: Each time Bob makes a measurement, he
chooses an initial condition for the reduced density matrix of
Alice. This explicitly involves the {\it ordinary\/}
reduction of the wave packet postulate. One can invent other
versions of the postulate, for example, a composition of a
nonlinear map $N$ with a projector $P$: $N^{-1}\circ P\circ N$,
as proposed by L\"ucke in his analysis of nonlinear gauge
transformations \cite{Lucke}. 

Alternatively, as in the Weinberg theory, one can get
the Gisin effect {\it without\/} the explicit use of the
projection postulate: The nonlinear generator of 
evolution must be basis dependent. If this is the case, the
2-particle solution is parametrized by the choice of the basis
in the Hilbert space. Assuming that Bob can change this basis we
allow him to change globally the form of the solution. 
But the rotation of the Stern-Gerlach device (modification of
the intarction term) is not yet a change
of basis in this sense. In the Weinberg theory the generator was
interpreted as an average energy. A change of basis was
equivalent to a different way of measuring the average. 
For this reason it was justified to say \cite{P} that
the effect was implied by the projection postulate although
explicit calculations were not referring to any projections
\cite{MCfpl} 
(similarly to the linear case, the projection is not included
in a non-stochastic Schr\"odinger dynamics). 

We can conclude that the Gisin-type reasoning has to be regarded
as ``unphysical", exactly in the same sense as the
Einstein-Podolsky-Rosen argument was unphysical according to
Bohr. 
The version of nonlinear theory we have discussed 
seems to be free of physical inconsistencies
and is not more nonlocal than the linear one. 

\bigskip
The idea of using the Haag-Bannier equation as a laboratory for 
testing the concepts of separbility was suggested to me by 
G.~A.~Goldin who also informed me about the results of his
joint work with G.~Svetlichny.
I am indebted to G.~A.~Goldin, 
H.-D.~Doebner, W.~L\"ucke, P.~Nattermann, and J.~Hennig for
valuable suggestions and critical comments, and to W. Puszkarz
who explained to me how to include into the framework 
the equations of Kostin \cite{Kostin} and Staruszkiewicz
\cite{Star,Pusz}. The work is a part
of the Polish-Flemish project 007 and was done during my stay at
the Arnold Sommerfeld Institute in Clausthal. I gratefully
acknowledge a support from DAAD.


\begin{references}
\bibitem{dm}Density matrices can, of course, evolve in a
nonlinear way.  
A spectacular example of a nonlinear evolution of a density
matrix is the interference in a Bose-Einstein condensate. It
is known that the interference fringes recently observed at MIT
\cite{MIT,MIT2} are well approximated by a nonlinear $|\psi(x)|^2$
Schr\"odinger equation. The atomic clouds in a trap
cannot be in a pure state because the atoms are highly
entangled with laser photons.
\bibitem{diss}For technical reasons 
I restrict my analysis to non-dissipative equations since
inclusion of imaginary terms in Hamiltonians does not allow for
an immediate application of the triple bracket results. This
does not mean that such an extension is impossible.
\bibitem{what}A general result guarantees that ${\rm Tr\,} (\rho^n)$ is
time independent and a Casimir invariant. This means, in
particular, that the triple-bracket dynamics preserves purity of
states \cite{MCpla,MCMM}.
\bibitem{problem}This appears to be in conflict with the
analysis given in \cite{BBM} where, assuming $F=F(|\psi(x)|)$ and
weak separability, it was concluded that $F(|\psi(x)|)=b\ln |\psi(x)|$.
There is no contradiction, however. Take
$\Psi(x_1,x_2)=\psi(x_1)\phi(x_2)$. According to our general
scheme the 2-particle nonlinearity is 
\begin{eqnarray}
\textstyle
{}&{}&
F\big ([{\textstyle \int} dz |\psi(x_1)\phi(z)|^2]^{1/2}\big)
+
F\big ([{\textstyle \int} dz |\psi(z)\phi(x_2)|^2]^{1/2}\big)\nonumber\\
&{}&=
F(|\psi(x_1)|)+F(|\phi(x_2)|)\nonumber
\end{eqnarray}
as expected. In \cite{BBM} it was assumed that the two particle
nonlinearity is $F(|\Psi(x_1,x_2)|)$.
\end{references}
\end{document}